# The Data Big Bang and the Expanding Digital Universe: High-Dimensional, Complex and Massive Data Sets in an Inflationary Epoch

Meyer Z. Pesenson*, Isaac Z. Pesenson**, Bruce McCollum*
*California Institute of Technology, **Temple University

**Abstract**

Recent and forthcoming advances in instrumentation, and giant new surveys, are creating astronomical data sets that are not amenable to the methods of analysis familiar to astronomers. Traditional methods are often inadequate not merely because of the size in bytes of the data sets, but also because of the complexity of modern data sets. Mathematical limitations of familiar algorithms and techniques in dealing with such data sets create a critical need for *new paradigms* for the representation, analysis and scientific visualization (as opposed to illustrative visualization) of heterogeneous, multiresolution data across application domains. Some of the problems presented by the new data sets have been addressed by other disciplines such as applied mathematics, statistics and machine learning and have been utilized by other sciences such as space-based geosciences. Unfortunately, valuable results pertaining to these problems are mostly to be found only in publications outside of astronomy. Here we offer brief overviews of a number of concepts, techniques and developments, some "old" and some new. These are generally unknown to most of the astronomical community, but are vital to the analysis and visualization of complex datasets and images. In order for astronomers to take advantage of the richness and complexity of the new era of data, and to be able to identify, adopt, and apply new solutions, the astronomical community needs a certain degree of awareness and understanding of the new concepts. One of the goals of this paper is to help bridge the gap between applied mathematics, artificial intelligence and computer science on the one side and astronomy on the other.

## I. Introduction

Astronomy is undergoing a rapid, unprecedented and accelerating growth in both the amount and the intrinsic complexity of data. This results partly from past and future large sky surveys: the Sloan Digital Sky Survey (York et al. 2000), the Large Synoptic Survey Telescope (LSST) (Tyson et al. 2008; see also Szalay et al. 2002), ESA's GAIA mission (Perryman 2002), Pan-STARRS (Hodapp et al. 2004; Price et al. 2007), the Palomar Transient Factory (Law et al. 2009), LAMOST (Zhao et al. 2006), and the Palomar-Quest Survey (Djorgovski et al. 2009). Smaller surveys and catalogs comprised of $\sim 10^3$-$10^4$ objects appear annually. The increasing availability of multiple-object spectrographs deployed at ground-based observatories enables observers to obtain spectra of hundreds of objects in a single exposure (Barden & Armandroff 1995; Bigelow et al. 1998; Lewis et al. 2002; Sharples et al. 2004; Wagner 2006; Freeman & Bland-Hawthorne 2008; Eto et al. 2004). Million-object spectrographs have been proposed and are undergoing design studies (e.g. Barden et al. 2004; Ditto & Ritter 2008). All together, and sometimes individually, such projects are creating for astronomy massive multitemporal and multispectral data sets comprised of images spanning multiple wavebands and including billions of objects. Furthermore, the Virtual Observatory (VO) is undertaking to combine existing historical data from all wavelengths into what will be, from the user's perspective, a single data set of gigantic size and unprecedented complexity (http://www.ivoa.net).

In addition, mathematically new (for astronomy) forms of data are starting to appear, such as those of the ESA Planck mission in which the cosmic microwave background (CMB) is characterized by a 2×2 matrix at each point in the sky (Zaldarriaga & Seljak 1996).

Other great challenges arise from the so-called three-dimensional (3D) reconstructions. For example, a very important and difficult problem of solar astrophysics is 3D reconstruction of coronal mass ejections (Frazin, et al. 2009). The resulting reconstruction problem cannot be solved via classical methods and must be addressed by more modern image processing methods like compressed sensing (see section V). Another example of tomographic reconstruction is coming from the neutral hydrogen mapping using the redshifted 21 cm line that has recently emerged as a promising cosmological probe. A number of radio telescopes are currently being proposed, planned or constructed to observe the redshifted 21 cm hydrogen line from the Epoch of Reionization (e.g. Fast Fourier Transform Telescope, Tegmark & Zaldarriaga 2009).

The richness and complexity of new data sets will provide astronomy with a wealth of information and most of the research progress expected from such sets inherently rests in their enormity and complexity. In order to take full advantage of immense multispectral, multitemporal data sets, their analysis should be **automated, or at least semi-automated,** and should consist of the following steps: **detection, characterization** and **classification** of various features of interest, followed, if necessary, by **automated decision-making** and possibly by **intelligent automatic alerts** (e.g. for follow-up observations or data quality problems). Moreover, a **real-time processing** may be required. Complex data also call for scientific visualization rather than ordinary illustrative visualization. By <u>scientific visualization</u> we mean visualization that <u>does not simply reproduce visible things, but makes the things visible</u>, thus enabling extraction of meaningful patterns from multiparametric data sets and ultimately facilitating analysis.

All this requires the development and adaption of modern methods for data representation, analysis and visualization. Methods now standard in astronomy are often ineffective because the data sets are too large and too complex for existing tools to be scaled in a straightforward way from handling several parameters up to dozens or more. There are also important limitations inherent in the mathematical algorithms familiar to astronomers. Scientific visualization, dimensionality reduction, and non-parametric methods in general are among the least-advanced categories of tools in astronomy because until this century there have not been data sets requiring new approaches related to those aspects of the data. All this not only creates a critical need for new sophisticated tools, but moreover urgently calls for *new paradigms* for the analysis, visualization and organization of heterogeneous, multiresolution data across application domains.

The astronomical community is becoming increasingly aware of the fact that new and advanced methods of applied mathematics, statistics and information technology should be developed and utilized. Three of the State of the Profession Position Papers submitted to the Astronomy and Astrophysics Decadal Survey strongly emphasized astronomy's need for new computational and mathematical tools in the coming decade (Loredo et al. 2009; Borne et al. 2009; Ferguson et al. 2009) (see: http://sites.nationalacademies.org/bpa/BPA_049492). A variety of individual and organizational initiatives and research projects have been proposed or are underway to at least partially address this looming challenge by developing or implementing a new generation of methods of data representation, processing and visualization.

However, there is a large communication gap between astronomy and other fields where adequate solutions exist or are being developed: applied mathematics, statistics and artificial intelligence.

The principal objectives of this paper are twofold. First, we wish to bring attention to some specific needs for new data analysis techniques. Second, we describe some innovative approaches and solutions to some of these problems and give examples of novel tools that permit effective analysis, representation, and visualization of the new multispectral/multitemporal data sets that offer enormous richness if they are mined with the appropriate tools. The extensive amount of relevant work already accomplished in disciplines outside of astronomy does not allow us to offer a complete review of all aspects of these complex topics and problems, but we have selected a number of important examples.

The structure of the paper is as follows. In part II we discuss challenges related to semi-automated processing of low-dimensional images and describe briefly our framework for advanced astronomical image processing. Part III describes problems posed by complex astronomical data sets; part IV addresses some of these problems and presents some innovative methods for nonlinear data dimension reduction, data representation and sampling on graphs and manifolds; it also describes briefly a new unifying method for image segmentation and information visualization that is based on physical intuition derived from synchronization of nonlinear oscillations; part V reviews some recent approaches to the challenges posed by high-dimensional, complex data sets and their importance for cosmological and astrophysical problems; in particular, we briefly discuss in this part a new tool called *needlets* for processing data that is laying on the sphere, applications of needlets to the analysis of CMB data, and a generalization of the wavelet-like transforms to Riemannian manifolds. The paper ends with a conclusion.

## II. Framework for Processing and Visualizing Astronomical Images

Vast data sets demand automated or semi-automated image processing and quality assessment of the processed images. Indeed, the sheer number of observed objects awaiting analysis makes obvious the need for sophisticated automation of object detection, characterization and classification. Adapting recent advances of computer vision and image processing for astronomy and designing and implementing an image processing framework (see below) that would utilize these continuing achievements, remains, however, a major challenge.

Developers and users alike have realized that there is more to creating an application than simple programming. The objective of creating a flexible (see below) application is customarily achieved by exploiting the object-oriented paradigm (Grand 1998; Booch et al. 2005). In computer science such systems are called frameworks (Fayad et al. 1999). Frameworks are flexible, and users can extend frameworks' capabilities by installing plug-ins without having to rewrite the basic code. An application framework consists of computational (processing) modules, data, and an interactive interface. Frameworks are extendable (i.e., they can easily adopt software products to changes of specification) and reusable (the software elements are able to serve for construction of many different new applications).

The keystone elements of a system that unifies a wide range of methods for astronomical image processing should be computational and visualization modules. Such a platform-independent framework with an integrated environment was described in Pesenson, Roby & McCollum (2008). It provides many common visualization features plus the ability to combine these images (overlays, etc.) in new and unique ways. The framework has a very intuitive GUI that allows the user to set up a custom pipeline interactively and to process images singly or in a batch mode. The final products, as well as the results of each step, are viewable with the framework. It also provides access to different data archives and can easily incorporate custom modules written in any programming language. Figures 1-4 give a few self-explanatory examples that demonstrate some of the framework's functionality.

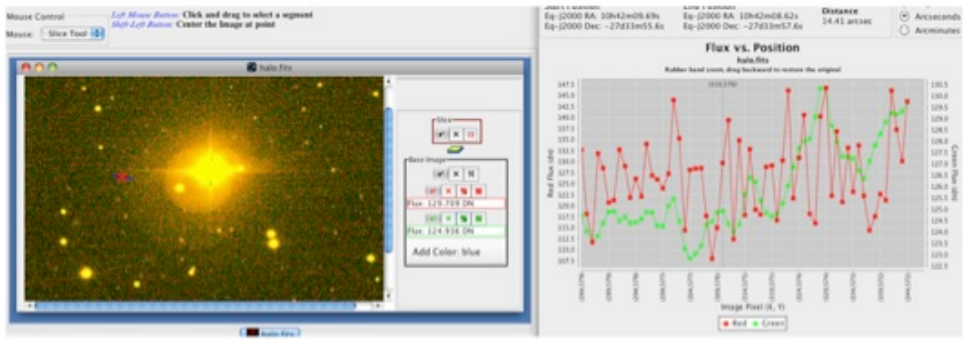

Figure 1. ESO-La Silla; courtesy of A. Grado, INAF-Osservatorio Astronomico di Capodimonte. Left: overlaid pre- and post-processed images; the red cross shows approximately the edge of the diffuse halo. Right: Flux-cut through the overlaid pre- and post-processed images (red and green respectively); the vertical grey line (close to the center of the plot) corresponds to the red cross in the plot on the left; after the preprocessing the average level "outside" of the halo is lower than the average level 'inside', thus enabling a better automated separation of the diffuse halo from the background.

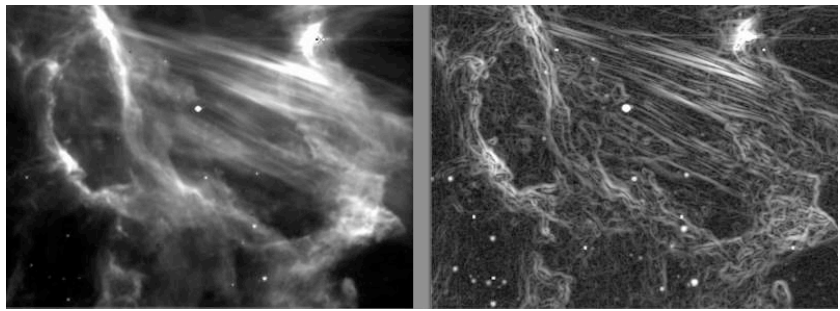

Figure 2. Morphology unveiling (Pesenson, Roby & McCollum 2008). Left: IC 405, *Spitzer* IRAC camera 8.0 μm image (France et al. 2007); filaments and a bow shock near HD34078 (see the red cross "x" in Figure 3, left). The brightness is proportional to the flux (logarithmic scale). Right: Brightness is proportional to the module of the gradient of flux (logarithmic scale); this way of looking at astronomical images facilitates analysis of nebular morphology, outflows, jets, embedded sources, and shock fronts.

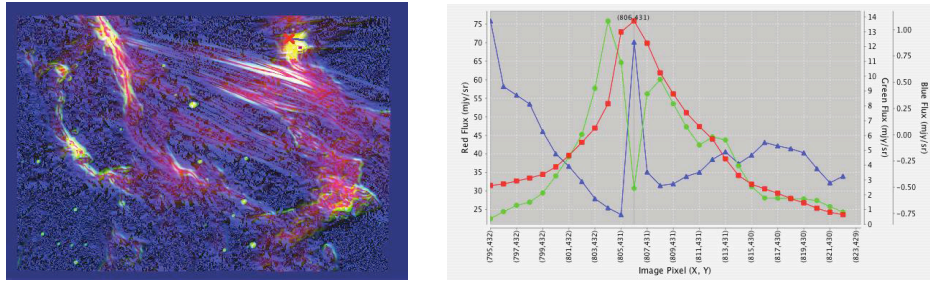

Figure 3. Overlaid pre- and postprocessed images (Pesenson, Roby & McCollum 2008). Left: Three overlaid images of IC 405: *Spitzer* IRAC camera 8.0 μm image, the module of its gradient, and the angle of its gradient (red, green, and blue respectively). The line through the red cross "x" indicates the crosscut. Right: Three profiles for the crosscut through ''x''. The crosshair goes through the ridge point and a local minimum of the module of the gradient. Note the different scales on the left and right vertical axes.

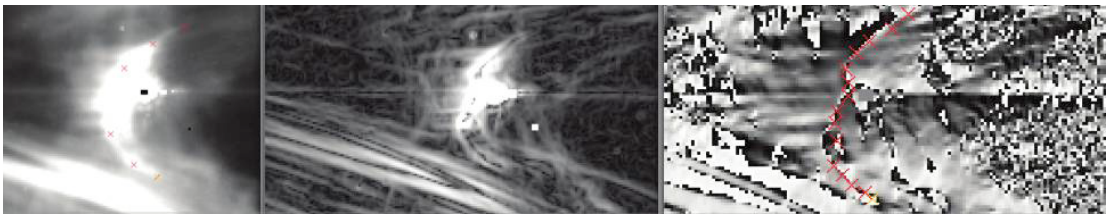

Figure 4. Three different visualizations of the bow shock structure (Pesenson, Roby & McCollum 2008). Left: IC 405, *Spitzer* IRAC camera 8.0 μm image, the bow shock (red crosses) near HD 34078. The brightness is proportional to the flux (logarithmic scale). Middle: The bow shock area where the brightness is proportional to the module of the gradient of flux (logarithmic scale). Right: The bow shock area where the front of the bow shock is immediately apparent as a curve of 1 pixel width (red crosses); the brightness is proportional to the angle of the gradient of flux (logarithmic scale).

   The framework deals primarily with image smoothing (regularization) and segmentation. These are fundamental first steps for the detection and characterization of image elements or objects and as such they play principal roles in the realization of automated computer vision applications. Image segmentation can roughly be described as the process that subdivides an image into its constituent parts (objects) and extracts those parts of interest. Since the inception of image segmentation in the 1960s, a large number of techniques and algorithms for image segmentation have been developed.

   Lately, due to revolutionary advances in instrumentation, the complexity of images has changed significantly: the extension of grey level images to multi- and hyperspectral images, from 2-D images to 3-D, from still images to sequences of images, tensor-valued images (polarization data), etc. Some modern, cutting-edge methods for image processing have been developed lately, and are being developed today, by information scientists outside of astronomy. The substantial progress in this direction made by the image processing and computer vision communities (Bovik 2005; Paragios et al. 2006), has found multiple applications in physics, technology, and bio-/medical sciences. Unfortunately, for the most part these advances have not
5

yet been utilized by the astronomical community.

Multiscale image representation and enhancement are such approaches. They have become important parts of computer vision systems and modern image processing. The multiscale approach has proven to be especially useful for image segmentation for feature and artifact detection. It enables a reliable search for objects of widely different morphologies, such as faint point sources, diffuse supernova remnants, clusters of galaxies, undesired data artifacts, as well as unusual objects needing detailed inspection by a scientist. It is well known that in astronomical images one often sees both point sources and extended objects such as galaxies embedded in extended emission (see, for example, Figs. 2, 6, 7). Because of the issue of robustness with respect to noise (section IV), a careful preprocessing is required before one can safely apply image segmentation or dimension reduction. So, an adequate way of preprocessing is what needs to be addressed first.

One effective approach to denoising is based on partial differential equations and may be seen as the local adaptive smoothing of an image along defined directions that depend on local intensities. One wants to smooth an image while preserving its features by performing a local smoothing mostly along directions of the edges while avoiding smoothing orthogonally to these edges. Many regularization schemes have been developed so far for the case of simple two-dimensional scalar images. An extension of these algorithms to vector-valued images is not straightforward. For a gray-scale image, the gradient is always perpendicular to the level set objects of the image; however, in the multi-channel case, this quality does not hold. Applying nonlinear diffusion to each channel or spectral band separately is one possible way of processing multi- and hyperspectral cubes; however, it does not take advantage of the richness of the multi/hyperspectral data. Moreover, if the edge detector acts only on one channel, it may lead to an undesirable effect, such as color blurring, where edges in multicolor images are blurred due to the different local geometry in each channel. Hence, a coupling between image channels should appear in the equations through the local vector geometry. We achieve this by implementing a nonlinear diffusion on a weighted graph, thus generalizing the approach adopted by Pesenson, Roby & McCollum (2008). The governing equation is linear, but the nonlinearity enters through the weights assigned to the graph edges. This algorithm respects the local smoothing geometry and thus serves well as a preprocessing step required for dimension reduction (section IV).

Our framework handles two-dimensional scalar images and paves the way to the semi- and automated image processing and image quality assessment. However, the ability to extract useful knowledge from high-dimensional data sets is becoming more and more important (section III). It is closely related to finding complicated structural patterns of interrelated connections and heavily depends on the ability to reduce dimensions of the raw data. This problem constitutes a big challenge for the scientific and technology communities. To extend the framework's functionality to high-dimensional images and data sets we have been developing novel, practical algorithms for dimension reduction. In the next section we describe some of the challenges presented by modern, high-dimensional data sets.

### III. Complex, Massive Data Sets

Astronomy has long found the use of multiple data dimensions to be crucial aids to progress. For example, surveys of H-alpha sources when cross-correlated with spectral types guided astronomy to discover interesting new types of objects such as Herbig AeBe stars (Herbig 1960). Cross-comparison of brightnesses, redshifts, and optical object morphologies led to the discovery of quasars. X-ray emission proved to be a highly efficient method of identifying pre-main sequence objects in a large field of view (e.g. review by Feigelson & Montmerle 1999). The fact that the same object (SS 433) was noticed to exhibit H-alpha emission and strong radio and optical variability led astronomers to investigate further and thus discover the existence of microquasars (Clark & Murdin 1978; Feldman et al. 1978; Margon et al. 1979). Other discoveries from the use of multidimensional data using just a few dimensions include such things as Be X-ray binaries and soft gamma ray repeaters.

Simple color-magnitude diagrams are another traditional tool taking advantage of multiple data dimensions (e.g. cataloging YSO candidates from Spitzer survey data, Whitney et al. (2008)). In addition, it is well known that simple color-color plots using four colors provide ways to efficiently and fairly reliably classify into physical types a large number of objects (IRAS data, e.g. van der Veen et al. 1989, van der Veen & Habing 1990; 2MASS data, e.g. Finlator et al. 2000; MSX data, e.g. Egan et al. 2001, Ortiz et al. 2005; *Spitzer* data, e.g. Ramirez et al. 2008) as well as to discover interesting new objects as outliers (e.g. Luminous Red Novae, Munari et al. 2002, Bond 2003). The meaning and usefulness of more complicated multiwavelength cross-correlations across widely-separated wavelength domains remain a nontrivial but fruitful challenge still being explored (for example, microquasars identified by comparing radio, IR, and X-ray properties (Combi et al. 2008)).

As multiwavelength data have become available for huge numbers of objects in the past few decades, the number of data dimensions for a typical object has grown beyond what can be visualized and studied using classical color-color plots and correlations using only a few dimensions. For example, Egan et al. (2001) write - "six-dimensional data are difficult to represent; in addition, it is not clear that all of the colors yield completely independent information" (see Figure 5).

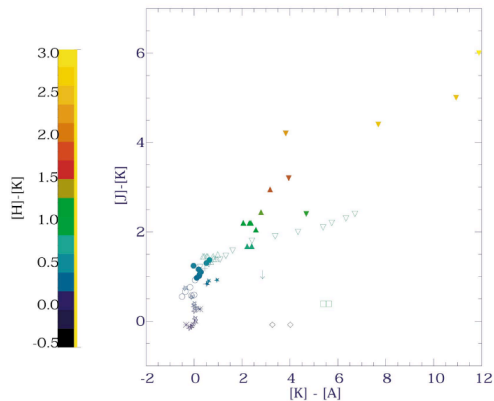

Figure 5. Color-color plots are a familiar form of scientific visualization. This example (from Egan et al. 2001) shows, simultaneously, infrared color-color information for several different categories of objects. Different object classes are represented by different symbols or different hues. Thus several data dimensions (the four IR colors and the several classes of object) are represented.

While the above figure successfully permits one to identify meaningful patterns, it is easily imagined that attempting to display a much larger number of data dimensions by simply using many more colors and symbols overlaid on the same plot and same region of parameter space would not be a useful form of scientific visualization. Also, two-dimensional plots are obviously not useful for expressing relationships among more than four colors. Effective scientific visualization of a large number of data dimensions requires new techniques.

The catalog information available for a great many objects already includes such information as magnitude, several optical and IR colors, metallicities, spectral types, and so on. A galaxy catalog would of course include other parameters such as morphology and redshift. When spectra are considered and compared in detail, the huge number of emission and absorption features obviously compounds the problem vastly. Still more complexity is added when one attempts to correlate a large grid of models with a large data set having many dimensions (e.g. the YSO analysis by Whitney et al. 2008) in order to create feedback for models based on statistically significant samples rather than on a few putative prototypes.

Other examples of high-dimensional data include, but are not limited to, multiparametric data sets (e.g. the manifold of galaxies: Brosche, 1973; Djorgovski & Davis, 1987), multitemporal, multispectral and hyperspectral data sets and images, and high-resolution simulations on massively parallel computers. The magnitude of the computational challenge for pattern recognition and classification algorithms is suggested by the fact that the VO will contain billions of data points each having hundreds of dimensions (Djorgovski et al. 2003).

All these examples clearly demonstrate that the automated and semi-automated processing required by the unprecedented and accelerating growth in both the amount and the complexity of astronomical data demands new ways of information representation. In the next section we discuss such approaches and describe some of the original algorithms we have developed in the course of this ongoing work.

**IV. Dimensionality Reduction and Sampling for Complex and Large Data Sets**

Because approaches to complex data require advanced mathematics, astronomers who wish to take advantage of them will need at least some basic knowledge of new, unfamiliar mathematical concepts and terminology. The full practical adoption of such methods requires interdisciplinary scientists who understand the new approaches in depth and are interested in working with astronomers to adapt and apply the methods to astronomical data sets. This is already happening as part of some research. The first step for a "neophyte" astronomer, however, is to learn "what is out there" as a basis for further investigation and consideration of the utility of various methods. The goal of this section is to offer a very basic introduction and explanation of some of these unfamiliar concepts.

Machine learning (Haykin 2009) is becoming increasingly important in astronomy (see an extensive review by Ball & Brunner 2009; also Longo et al. 2004). The main objectives of machine learning are clustering (automatic identification of groups of similar objects), and classification (assigning labels to instances). However, high dimensionality complicates machine learning and can easily thwart the entire effort (Haykin 2009 and references there). It also becomes a formidable obstacle in computing numerical solutions in Bayesian statistics for models with more than a few parameters.

R. Bellman (1961) coined the term "curse of dimensionality", to describe how difficult it was to perform high-dimensional numerical integration. For example, 100 evenly-spaced sample points suffice to sample a unit interval with no more than 0.01 distance between points; an equivalent sampling of a 10-dimensional unit hypercube with a lattice with a spacing of 0.01 between adjacent points would require $10^{20}$ sample points. Thus, in some sense, the 10-dimensional hypercube can be said to be a factor of $10^{18}$ "larger" than the unit interval. *Obviously, this will make many computational tasks intractable for high-dimensional data sets.*

Euclidian spaces are usually used as models for traditional astronomical data types (scalars, arrays of scalars). Another *general manifestation* of high dimensionality is the fact that in a high-dimensional Euclidean space, volume expands far more rapidly with increasing diameter than it expands in lower-dimensional spaces. Indeed, if one compares the size of the unit ball with the unit cube as the dimension of the space increases, it turns out that the unit ball becomes an insignificant volume relative to that of the unit cube. Thus, in some sense, nearly all of the high-dimensional space is "far away" from the center. This is called the "empty space phenomenon" (Scott 1992; Matousek 2002): the high-dimensional unit space can be said to consist almost entirely of the "corners" of the hypercube, with almost no "middle". One more important example of the unexpected properties of high-dimensional Euclidean spaces is the following behavior of the Gaussian distribution in high dimensions: the radius of a hypersphere that contains 95% of the distribution grows as the dimensionality increases.

These sorts of unexpected mathematical properties demonstrate that in order to make practical the extraction of meaningful structures from multiparametric, high-dimensional data sets, a low-dimensional representation of data points is required. Dimension reduction (DR) is motivated by the fact that the more we are able to reduce the dimensionality of a data set, the more regularities (correlations) we have found in it and therefore, the more we have learned from the data. Data dimension reduction is an active branch of applied mathematics and statistics (Haykin 2009). It consists of methods for finding lower-dimensional representation of high-dimensional data, without losing a significant amount of information, by constructing a set of basis functions that capture patterns intrinsic to a particular state space. DR methods greatly increase computational efficiency of machine learning algorithms, improve statistical inference and enables effective scientific visualization and classification. From a large set of images obtained at multiple wavebands, effective dimension reduction provides a comprehensible, information-rich single image with minimal information loss and statistical details, unlike a simple coadding with arbitrary, empirical weights (see a simple four-wavelengths example in Fig. 6,7).

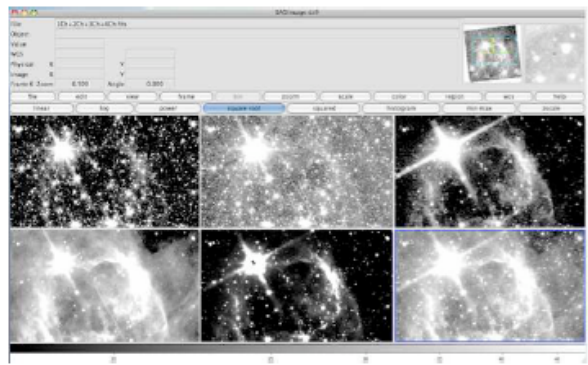

Figure 6. *Spitzer* IRAC camera 8.0 µm image, 1-4 channels. Upper row: channels 1-3. Lower row: channel 4; the result of applying dimension reduction; sum of the channels 1-4.

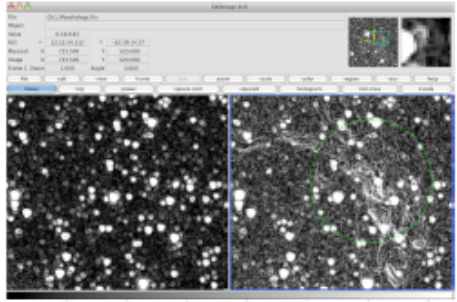

Figure 7. *Spitzer* IRAC camera 8.0 µm image, Left: channel 1, morphology unveiling following Pesenson, Roby & McCollum, 2008. Right: morphology unveiling performed on the reduced image from Fig. 6. It demonstrates that dimensionality reduction retains the important features from all channels. This is especially useful for dealing with large numbers of images.

Classical approaches to dimension reduction not unfamiliar to astronomy are principal components analysis (PCA) (Bijaoui 1974) and multidimensional scaling (Borg & Groenen 2005). Although first used in astronomy by Bijaoui (1974), PCA was not commonly used until the 1990's (Boroson & Green, 1992 and Francis et al. 1992; see also a general introduction to PCA by Francis & Wills, 1999). It has been used for such things as classification of galaxies and quasars (Francis et al. 1992; Connolly et al. 1995), photometric and spectroscopic redshift estimation (e.g. Glazebrook, Offer, & Deeley 1998; Yip et al. 2004), sky subtraction (Wild & Hewitt, 2005) and optical spectral indicators (Wild et al. 2007). PCA has been used for the simultaneous analysis of dozens of data parameters for each member of a sample of 44 active galactic nuclei (Kuraszkiewicz et al. 2009), which appears to be the largest number of parameters yet analyzed using PCA in an astronomical study.

PCA has a serious drawback in that it does not explicitly consider the structure of the manifold on which the data may possibly reside. In differential geometry an n-dimensional manifold is a metric space that on a small scale resembles the n-dimensional Euclidean space; thus a circle is a one-dimensional manifold, while a sphere is a two-dimensional manifold. PCA is intrinsically linear, so if data points form a nonlinear manifold, then obviously, there is no rotation & shift of the axis (this is what a linear transform like PCA provides) that can "unfold" such manifold. In other words, if the data are mainly confined to an almost linear low-dimensional subspace, then simple linear methods such as PCA can be used to discover the subspace and estimate its dimensionality. If, on the other hand, the data lie on (or near) a highly nonlinear low-dimensional submanifold, then linear methods are not effective in capturing the fine meaningful structures in the data. A blind application of linear methods may result in a complete misrepresentation of the data.

We have recently developed some advanced, original methods for performing nonlinear DR, which do not suffer from the limitations of PCA. In what follows, we briefly describe these methods. First, we introduce some more concepts and methods that have proved to be effective in the area of machine learning.

**Graphs.** In the context of data retrieval and processing, dimensionality reduction methods based on *graph theory* have proved to be very powerful. In mathematics and computer science, graph theory is the study of mathematical structures to model relations between objects (Gross & Yellen 2004). Graph theory has been successfully applied to a wide range of very different disciplines, from biology to social science, computing and physics. Among many applications, graphs and manifolds have been used to address mining databases, Internet search engines, computer graphics, computation reordering, image processing, etc. In what follows we discuss various data structure and algorithms that we have already developed and are developing.

The graph representation of structured data provides a fruitful model for the relational data mining process. A graph is a collection of nodes and links between them; the nodes represent data points and the weights of the links or edges indicate the strength of relationships. A graph in which each graph edge is replaced by a directed graph edge is called a directed graph, or diagraph (Gross & Yellen 2004). Diagraphs are used for context-sensitive browsing engines and for ranking hyper-linked documents for a given query.

The modern approach to multidimensional images or data sets is to approximate them by graphs or Riemannian manifolds. The first important, and very challenging, step is to convert such a data cloud to a weighted finite graph. The next important problem is the choice of "right" weights that should be assigned to edges of the constructed graph. The weight function describes a notion of "similarity" between the data points and as such strongly affects the analysis of the data. The weights should entirely be determined by application domain. The most obvious way to assign weights is to use a positive kernel like an exponential function whose exponent depends on the local Euclidean distance between data points and a subjectively chosen parameter called "bandwidth". (There are also other ways of assigning weights, which depend on more complex mathematics than we discuss in this article.)

Next, after constructing a weighted graph, one can introduce the corresponding combinatorial Laplace operator (Chung 1996; Gross & Yellen 2004). In machine learning, methods based on graph Laplacians have proved to be very powerful (Haykin 2009). The eigenfunctions and eigenvalues of the Laplacian form a basis, thus allowing one to develop a harmonic or Fourier analysis on graphs. By further developing harmonic analysis on graphs and manifolds (Pesenson 2004-2009; Pesenson & Pesenson 2009; Pesenson, Pesenson, & McCollum 2009), we have devised innovative algorithms for data compression and nonlinear data dimension reduction. These results enable one to overcome PCA's limitations for handling nonlinear data manifolds and also allow one to deal effectively with incomplete data (such as missing observations or partial sky coverage).

**Hypergraphs.** Most existing data mining and network analysis methods are limited to pairwise interactions. However, sets of astronomical objects usually exhibit multiple relationships, so restricting analysis to the dyadic (pairwise) relations leads to loss of important information and to missing discoveries. Triadic, tetradic or higher interactions offer great practical potential. This has led to the approach based on hypergraphs (e.g. Zhou, Huang & Scholkopf 2005). Hypergraphs constitute an important extension of graphs that allow edges to connect more than two vertices simultaneously. Thus hypergraphs provide a more comprehensive description of feature relations and structures. Hypergraphs furnish a much more adequate approach to real world data sets and allow one to deal with clustering and classifications using higher order relations. It has been shown that, in general, there does not exist a graph model that correctly represents the cut properties of the corresponding hypergraph (Ihler et al. 1993). Thus new mathematical methods are required to take advantage of the richness of information available in hypergraphs. In order to provide means for analysis of databases with multiple relationships, we are currently developing extensions of the original methods described in the previous subsection.

**Fractals.** Data sets having fractional dimensions ("fractals": Mandelbrot 1977; Schroeder 1991; Faloutsos 2003) have been suggested to represent many phenomena in astronomy. Examples include star formation on galactic scales (Elmegreen & Elmegreen 2001), the stellar initial mass function (Larson 1992), galaxy distributions in space (Einasto et al. 1991; Kurokawa, Morikawa & Mouri 2001), cloud distributions in active galactic nuclei (Bottorff & Ferland 2001), the cosmic microwave background (de Gouveia dal Pino et al. 1995), thermonuclear flame velocities in Type Ia supernovae (Ghezzi, de Gouveia dal Pino & Horvath 2004), and the mass spectrum of interstellar clouds (Elmegreen & Falgarone 1996). Such data sets exhibit a dimensionality which is often much lower than the dimension of the Euclidian space they are embedded into. For example, the distribution of galaxies in the universe has dimension $D \sim 1.23$. The difference between the two dimensions occurs because the fractal dimensionality is intrinsic. The intrinsic dimension of a graph reflects the intrinsic dimension of the sampled manifold.

Obviously, an important first step in practical dimensionality reduction is a good estimate of the intrinsic dimension of the data. Otherwise, DR is no more than a risky guess since one does not know to what extent the dimensionality can be reduced. To enable analysis of astronomical data sets that exhibit a fractal nature, we are currently developing a practical concept of spectral dimensionality, as well as original algorithms for sampling, compression and embedding fractal data.

**The Petascale Connection.** The approaches described above dealt with compact manifolds and finite graphs. However, massive data sets are more adequately described by non-compact manifolds and infinite graphs. In order to deal with extremely large data sets we extended dimension reduction to *non-compact* manifolds and *infinite* graphs. We are also working on generalizing the Fourier analysis to *non-compact* Riemannian manifolds and *infinite* quantum and combinatorial graphs, directed graphs, hypergraphs and some fractals. Implementation of these algorithms, and incorporation of them into the framework described in section II, will enable a more adequate analysis of massive data sets.

**Robustness of Dimension Reduction Algorithms with Respect to Noise.** Despite the important and appealing properties of the above mentioned dimension reduction algorithms, both linear and nonlinear approaches are sensitive to noise, outliers, and missing measurements (bad pixels, missing data values). Because noise and data imperfections may change the local structure of a manifold, locality preservation means that existing algorithms are topologically unstable and not robust against noise and outliers (Balasubramanian & Schwartz 2002).

This is obviously a serious drawback because astronomical data are always corrupted by noise. Budavari et al. (2009) presented a robust version of PCA and applied it to astronomical spectra. Their approach addressed the issues of outliers, missing information, large number of dimensions and the vast amount of data by combining elements of robust statistics and recursive algorithms that provide improved eigensystem estimates step by step. However, as it was mentioned earlier, the PCA method is intrinsically linear and cannot be applied to data points residing on a nonlinear manifold.

Thus the practical usage of dimension reduction demands careful improvement of signal-to-noise ratio without smearing essential features. Implementing a nonlinear diffusion on weighted graphs (section II) enables us to apply dimension reduction to noisy data sets as well.

**Image Segmentation, Smoothing and Information Visualization.** In what follows we briefly describe a new, original unifying approach to segmentation of images in particular and pattern recognition and information visualization in general. Image segmentation (see section II) plays a principal role in the realization of automated computer processing, as a previous stage for the recognition of different image elements or objects.

Solutions to the manifold learning problem can be based on intuition derived from physics. A good example of this is the approach of Horn & Gottlieb (2002) that is based on the Schrödinger operator, where they constructed an approximate eigenfunction, and computed its corresponding potential. The clusters were then defined by the minima of the potential.

We developed (Pesenson & Pesenson 2010) an alternative that is also based on physical intuition, this one being derived from synchronization of nonlinear oscillations (Kuramoto 1984; Pikovsky et al. 2001). Approximating a multidimensional image or a data set by a graph and associating a nonlinear dynamical system with each node enables us to unify the three seemingly unrelated tasks: image segmentation, unsupervised learning and data visualization. Pattern recognition may benefit significantly from new methods of visualization and representation of data, since they may uncover important relationships in multidimensional data sets. Bringing out patterns by setting data points into oscillatory motion is a very effective way of visualizing data (Ware & Bobrow 2004). Our method is a feature-based, multilevel algorithm that finds patterns by employing a nonlinear oscillatory motion. At the same time, the oscillatory motion reveals to the eye patterns by making them oscillate coherently with a frequency different from the rest of the graph. Patterns are detected recursively inside the graph, and the found features are either collapsed into single nodes, forming a hierarchy of patterns or can be zoomed in and studied individually. This method can be described as a pre-computed animation and it enables both qualitative (by eye) and quantitative discovery of correlations (see Figure 8).

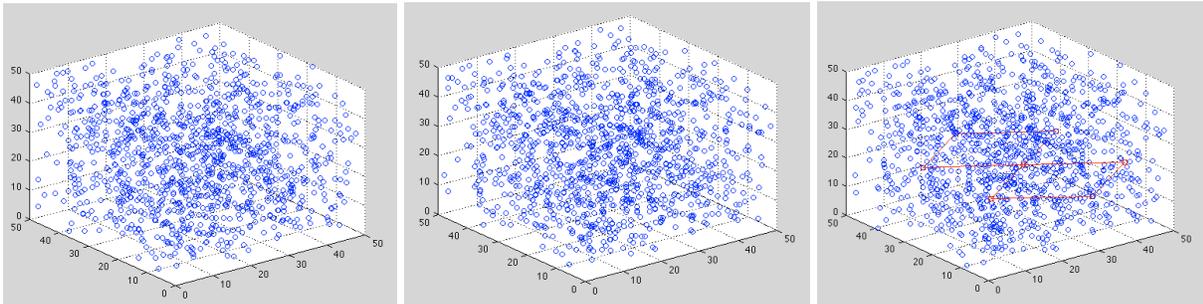

Figure 8. Testing the algorithm on synthetic data. A simulated three-dimensional set of a thousand uniformly distributed random points with a double-diamond pattern created by assigning large weights to the edges connecting the points in the pattern (while the rest of the weights are negligible). Left and middle: two screen shots from a running animation - each point in the set oscillates (in this case in three dimensions) with its own, random frequency. Right: synchronization made the points that are connected with high-weight edges oscillate in-phase thus allowing to reveal the pattern visually (to avoid extra clutter, the edges are not displayed in the animation), or by automatically selecting in-phase oscillating points and highlighting the pattern in red.

The approaches presented in this section enable interpolation, smoothing and immersions of various complex (dozens or hundreds of useful parameters associated with each astronomical object) and large data sets into lower-dimension Euclidean spaces. Classification in lower dimensional space can be done more reliably than in high dimensions. Thus, DR can be significantly beneficial as a pre-processing step for many existing astronomical packages, such as for example, the popular source extractor SExtractor (Bertin & Arnouts 1996). We also described a new method for pattern recognition (clustering) and visualization of multidimensional data. Altogether, these approaches provide important generalizations of the tools for spectral clustering and dimensionality reduction, and enable more adequate representation, effective data retrieval and analysis of complex, modern astronomical data sets.

## V. Some Recent Approaches to the Challenges of Data Intensive Astronomy

Data intensive astrophysics requires an interdisciplinary approach that will include elements of applied mathematics (Goodman & O'Rourke 2004; Gross & Yellen 2004), modern statistical methods (Feigelson & Babu 2003: Silk 2003), machine learning (Haykin 2009), computer vision (Paragios et al. 2006) and image processing (Bovik 2005). The breadth and complexity of the work relevant to modern astronomical data challenges is large, and does not permit a full treatment here. However, it is valuable to briefly mention a number of other important problems, approaches and efforts that have been pursued.

The problem of processing data that lay on a manifold is very important for cosmological data analysis. The standard, powerful data analysis package HealPix (Gorski et al. 2005) processes data on a two-dimensional manifold – the sphere. The concept of needlets (second generation spherical wavelets) has recently attracted a lot of attention in the cosmological literature. The first application of needlets to cosmological data was provided by Pietrobon et al. (2006). They analyzed cross-correlation of Wilkinson Microwave Anisotropy Probe (WMAP) CMB data with NRAO VLA Sky Survey (NVSS) radio galaxy data. The approach based on needlets enabled more accurate statistical results related to the dark energy component. The investigation of needlets from the probabilistic viewpoint and their relevance for the statistical analysis of random fields on the sphere was done for the first time by Baldi et al. (2006). A thorough presentation of the spherical needlets for CMB data analysis is given by Marinucci et al. (2008). Various issues related to CMB, such as spectrum estimation, detection of features and anisotropies, mapmaking were addressed by Fay et al. (2008), Pietrobon et al. (2008) (see also references there). The CMB models are best analyzed in the frequency domain, where the behavior at different multipoles can be investigated separately; on the other hand, such problems as missing observations or partial sky coverage make impossible the evaluation of spherical harmonic transforms. The needlets allow for a very efficient simple reconstruction formula that enables one to perform frequency analysis by using only partial information about data and providing means for handling masked data. Besides, needlets exemplify other important properties that are not generally shared by other spherical wavelet constructions: they do not rely on any kind of tangent plane approximation; they have good localization properties in both pixel and harmonic space; and needlet coefficients are asymptotically uncorrelated at any fixed angular distance (which makes their use in statistical procedures very promising). All these unique properties make needlets a very valuable tool in various areas of CMB data analysis. Recently, Geller & Marinucci (2008) introduced spin needlets (a new form of (spin) wavelets) as a tool for the analysis of spin random fields. Geller et al. (2008) adopted the spin needlet approach for the analysis of CMB polarization measurements. Lately Geller & Mayeli (2009) constructed continuous wavelets and nearly tight frames (needlets) on compact manifolds.

The essential part of the general analysis based on the wavelet-like constructions is sampling of bandlimited functions. The mathematical foundations of sampling on arbitrary Riemannian manifolds of bounded geometry were laid down by Pesenson (1995, 2000). Recently Geller & Pesenson (2010), constructed bandlimited and highly concentrated tight frames on compact homogeneous manifolds. These results can be considered as an extension of the wavelet-like transforms to Riemannian manifolds, thus enabling one to reconstruct information from incomplete data defined on an arbitrary manifold.

Multiscale data analysis has proved to be a very powerful tool in many fields. Applications of multiscale image analysis to astronomy were pioneered by Starck, Murtagh & Bijaoui (1998) and Starck & Murtagh (2002).

Spectral methods and diffusion maps have recently emerged as effective approaches to nonlinear dimensionality reduction (Belkin & Nyogi 2005; Coifman & Lafon 2006; Lafon & Lee 2006) The diffusion maps approach has successfully been applied to analysis of astronomical spectra by Richards at al. 2009.

Manifold learning may be seen as a DR procedure aiming at capturing the degrees of freedom and structures (clusters, patterns) within high-dimensional data. Manifold learning nonlinear algorithms such as isometric mapping (ISOMAP) by Tenenbaum et al. (2000) and local linear embedding (LLE) by Roweis & Saul (2000) project high-dimensional manifold data into a low-dimensional space by preserving the local geometric features. Ball & Brunner (2009) provide a very broad review of the current state of machine learning and data mining in astronomy.

The AstroNeural collaboration group (Longo et al. 2004) implemented tools based on neural networks, fuzzy-C sets and genetic algorithms, and applied them to perform complex tasks such as unsupervised and supervised clustering and time series analysis. D'Abrusco et al. (2007) present a supervised neural network approach to the determination of photometric redshifts. D'Abrusco et al. (2009) describe a method for the photometric selection of candidate quasars in multiband surveys based on the probabilistic principal surfaces. Pasian et al. (2007) give a general review of the development of tools to be subsequently used within the international VO.

Comparato et al. (2007) show how advanced visualization tools can help the researcher in investigating and extracting information from data. Their focus is on VisIVO, a new open‐source graphics application that blends high performance multidimensional visualization techniques and up‐to‐date technologies to cooperate with other applications and to access remote, distributed data archives.

Draper et al. (2007) discuss the GAIA application for analyzing astronomical image and show how the PLASTIC protocol has been used to inter-operate with VO enabled applications.

The Center for Astrostatistics (CASt) at Pennsylvania State University provides a wealth of resources (codes, data, tutorials, programs, etc.) related to challenges in statistical treatments of astrophysical data: http://astrostatistics.psu.edu/

A large amount of practical and up-to-date information (texts, tutorials, preprints, software, etc.) related to Bayesian inference in astronomy and other fields is provided by T. Loredo on the website BIPS (Bayesian Inference for the Physical Sciences (BIPS) at http://www.astro.cornell.edu/staff/loredo/bayes/index.html

The International Computational Astrostatistics (InCA) Group at Carnegie Mellon University develops and applies new statistical methods to inference problems in astronomy and cosmology, with an emphasis on computational nonparametric approaches (see for details http://www.stat.cmu.edu/~inca/index.html).

The AstroMed project at Harvard University's IIC is dedicated to the application of medical image visualization to 3D astronomical data (Borkin et al. 2007).

Compressed sensing and the use of sparse representations offer another promising new approach. Traditionally it has been considered unavoidable that any signal must be sampled at a rate of at least twice its highest frequency in order to be represented without errors. However, a technique called compressed sensing that permits error-free sampling at a lower rate has been the subject of much recent research. It has great promise for new ways to compress imaging without significant loss of information, thus ameliorating analysis limitations deriving from limitations of computing resources. The power of compressed sensing was strikingly illustrated when an object was successfully imaged in some detail by a camera composed of a *single pixel* (Chan et al. 2008). Bobin et al. (2008) discuss recent advances in signal processing that use sparse representations. Relatively nontechnical introductions to compressed sensing may be found in Candes & Wakin (2008) and Romberg (2008). The first astronomical results in the literature based on the use of compressed sensing are that of Wiaux et al. (2009).

Various data types together with methods used for their representation are briefly summarized in Table 1. The table is not comprehensive, but it provides a quick overview of what was discussed above.

| Data Types | Some Astronomical Applications | Traditional Approaches to Data Representation & Processing | Advanced Approaches to Data Representation & Processing |
|---|---|---|---|
| Vector Data | 1. Multiwavelength observations. 2. Multitemporal observations. 3. VO 4. Spectra. | 1. Linear dimension reduction: PCA and its modifications. | 1. Spectral methods, eigenmaps, diffusion maps, LLE, ISOMAP. 2. Sampling on graphs. 3. Methods based on nonlinear dynamics. 4. Neural networks, fuzzy-C sets. 5. Genetic algorithms. 6. Scientific visualization. 7. Compressed sensing. |
| Manifold - Valued and/or Manifold - Defined | 1. Polarization measurements (CMB). 2. Gravitational lensing. 3. Solar astrophysics. | 1. Various sampling distributions on the sphere. | 1. Healpix (data on 2D sphere). 2. Needlets. 3. Sampling on manifolds. 4. Scientific visualization. |

Table 1. Examples of complex data types and some of the methods for their representation and processing.

# Conclusion

Extremely large data sets, as well as the analysis of hundreds of objects each having a large number of data dimensions, present astronomy with unprecedented challenges. The challenges are not only about database sizes in themselves, but about how intelligently one organizes, analyzes, and navigates through the databases, and about the limitations of existing data analysis approaches familiar to astronomy. The answers to these challenges are not trivial, and for the most part lie in complex fields of research well outside the training and expertise of almost all astronomers. Fortunately, other disciplines such as imaging science and earth sciences have for many years been grappling with the same sorts of problems. Fruitful interdisciplinary work has already become a regular feature of research in those other disciplines, and has resulted in applications of crucial value to other sciences seeking to take advantage of complex, giant data sets in their respective fields. This work has brought about many helpful applications and promising paths for further progress that potentially have significant value to astronomy.

Multidimensional image processing, image fusion (combining information from multiple sensors in order to create a composite enhanced image) and dimension reduction (finding lower-dimensional representation of high-dimensional data) are effective approaches to tasks that are crucial to multitemporal, multiwavelength astronomy: study of transients, large-scale digital sky surveys, archival research, etc. These methods greatly increase computational efficiency of machine learning algorithms and improve statistical inference, thus facilitating automated feature selection, data segmentation, classification and effective scientific visualization (as opposed to illustrative visualization). Dimensionally reduced images also offer an enormous savings in storage space and database-transmission bandwidth for the user, without significant loss of information, if appropriate methods are used.

To effectively use the large, complex data sets being created in $21^{st}$ Century astronomy, significant interdisciplinary communication and collaboration between astronomers and experts in the disciplines of applied mathematics, statistics, computer science and artificial intelligence will be essential. The concepts and approaches described in this paper are among the first steps in such a broad, long-term interdisciplinary effort that will help bridge the communication gap between astronomy and other disciplines. These concepts, and the approaches derived from them, will help to provide practical ways of analysis and visualization of the increasingly large and complex data sets. Such sophisticated new methods will also help to pave the way for effective automated analysis and processing of giant, complex data sets.

# Acknowledgement

The authors would like to thank Alanna Connors for stimulating discussions and useful suggestions. One of us (MP) would like to thank Michael Werner for support. We would also like to thank the referee for constructive suggestions that led to substantial improvements of the article. This work was carried out with funding from the National Geospatial-Intelligence Agency University Research Initiative (NURI), grant HM1582-08-1-0019, and support from NASA to the California Institute of Technology and the Jet Propulsion Laboratory.